\documentclass[
    ,final            
  ]
  {aipproc}

\layoutstyle{6x9}
\begin{document}
\title{Azimuthal Correlations in p-p collisions}


\author{Eleazar Cuautle}{
  address={Instituto de Ciencias Nucleares, UNAM, A. P. 70543, 04510 Mexico City, Mexico.}
}
\author{Isabel Dom\'inguez }{
  address={Instituto de Ciencias Nucleares, UNAM, A. P. 70543, 04510 Mexico City, Mexico.}
}
\author{Guy Pai\'c}{
  address={Instituto de Ciencias Nucleares, UNAM, A. P. 70543, 04510 Mexico City, Mexico.}
}

\begin{abstract}

We report the analysis of experimental azimuthal correlations measured by STAR
in p-p collisions at $\sqrt{s_{NN}}$ = 200 GeV. We conclude that for a fit of
data using Pythia event generator we need to include two values of $k_{T}$.
\end{abstract}

\maketitle

\section{Introduction}
Jets are produced by the hard scattering of two partons. Two scattered partons
propagate nearly back-to-back in azimuth from the collision point and 
fragment into jet-like spray of final state particles (The schematic view
is in Figure 1). These particles have a transverse momentum $j_{T}$ 
with respect to the parent partons, with component $j_{Tz}$ projected onto 
the azimuthal plane. The magnitude of $j_{Tz}$ measured at lower energies 
has been found to be $\sqrt{s_{NN}}$ and $p_{T}$ independent. 

In collinear partonic collisions, the two partons emerge with the same
magnitude of transverse momentum in opposite directions. However, the partons
carry the ``intrinsic'' transverse momentum $k_{T}$ before the collision. This
momentum affects the outgoing transverse momentum $p_{T}$, resulting in a
momentum imbalance (i.e. transverse momentum of one jet does not lie in the
plane determined by the transverse momentum of the second jet and the beam
axes) and consequently affects the back-to-back correlations of final high
$p_{T}$ hadrons \cite{R.P.Feynman}.

\begin{figure}[h!]
\includegraphics[height=0.4\textheight]{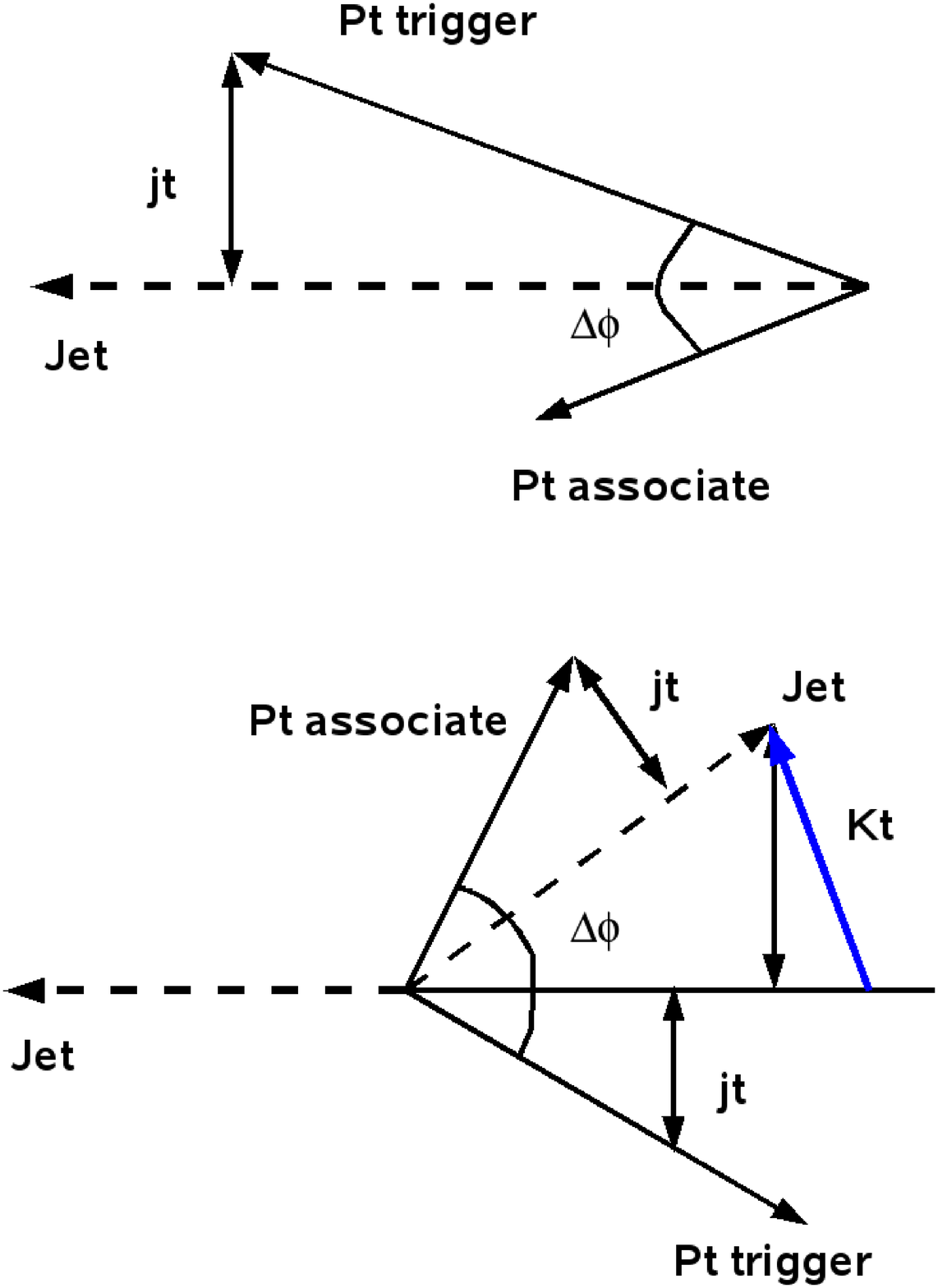}
\caption{Schematic view of a jet fragmentation, near-side jet (upper) and
  away-side jet (lower)}
\end{figure}

The back-to-back azimuthal correlations of high $p_{T}$ hadrons is written as

\begin{equation}
  C(\Delta\Phi)=\frac{1}{N_{trigger}}\int d \Delta\eta\frac{dN}{d 
  \Delta\Phi d\Delta\eta}
\end{equation}

\hspace{-0.4cm}where $\Delta\Phi$ and $\Delta\eta$ are, respectively, the
azimuthal angle and pseudorapidity between a trigger and their associated
particles.  The azimuthal correlation function displays a two-peak structure, 
where the width of the near-side peak is denoted by $\sigma_{N}$ and the width 
of the away-side peak is $\sigma_{A}$. The value of $\sigma_{N}$ carries
information on the fragmentation process only i.e. $j_{T}$. For particles with
average transverse momenta  $<p_{T, trigg}>$  and $<p_{T,  associate}>$ from
the same jet, the width of the near-side correlation,  $\sigma_{N}$, can be
related to $<j_{Tz}>$ as \cite{Rak:2001}:

\begin{equation}
  <j_{Tz}>=\frac{<p_{T,trigger}><p_{T,associate}>}{\sqrt{<p_{T,trigger}>^2+<p_{T,associate}>^2}} \sigma_{N}
\end{equation}

The width of the away-side peak  $\sigma_{A}$ contain the contribution of the 
intrinsic transverse momentum $k_{T}$. It has been characterized by a Gaussian
distribution \cite{Wang:1998ww}

\begin{equation}
 g(k_{T})=\frac{1}{2\pi\sigma^{2}}exp({-\frac{k_{T}^{2}}{2\sigma^{2}}})
\end{equation}

The azimuthal correlations are used extensively in heavy ions collisions to
understand the parton suppression mechanisms. We have concentrated in this work
to the simplest case i.e. p-p to understand the size and details of the peaks
in azimuthal correlations.

\section{$k_{T}$ contribution in the azimuthal correlations}

Correlation function was calculated choosing $<k_{T}^{2}>$=0, 1, 4 $
GeV^{2}/c^{2}$ at 200 GeV in a mid-rapidity region ($\mid\eta\mid <$ 0.7).
 Charged hadrons in 4 < $p_{T, trigg}$ < 6 GeV/c and in 
2 GeV $<$ p$_{T,assoc} <$ 4 GeV are defined to be trigger and
 associated particles respectively. In the actual calculation, we use PYTHIA
 6.325 \cite{Sostrand2001} in AliRoot \cite{ALICE} to simulate each hard scattering  
where a Gaussian distribution is assuming for $k_{T}$. 
The correlations functions were fitted  by the sum of two Gaussians, one for 
the near-side component (around $\Delta\Phi$ = 0 radians) and one for the 
away-side component (around $\Delta\Phi$ = $\pi$   radians) and a constant for
the uncorrelated pairs.

Figure 2 compares experimental data \cite{Adams:2003im} with different
$<k_{T}^{2}>$ simulations. In the four cases is observed that can not
reproduce experimental data. In order to reproduce the experimental data, we characterize the intrinsic momentum by two Gaussians distributions

\begin{figure}
  \includegraphics[height=.45\textheight]{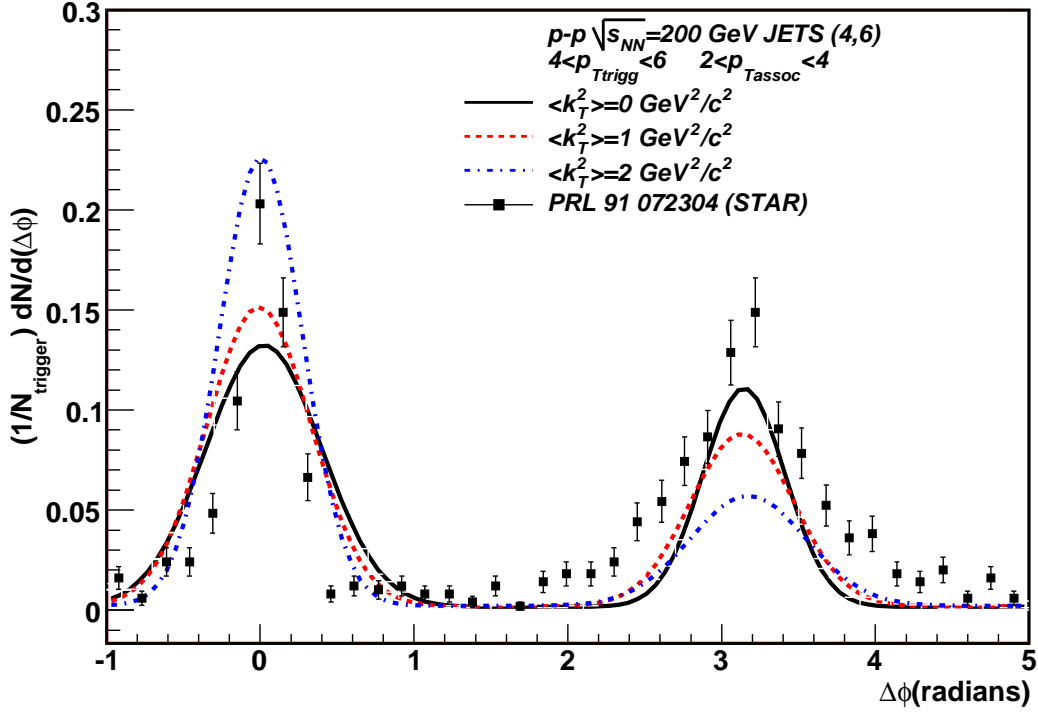}
  \caption{Azimuthal distributions for p+p collisions at $\sqrt{s_{NN}}$ = 200
    GeV, experimental data \cite{Adams:2003im} and
    simulations with different $<k_{T}^{2}>$ }
\end{figure}

\begin{equation}
  g(k_{T1},k_{T2})=\frac{1}{2\pi\sigma_{1}^{2}}exp({-\frac{k_{T1}^{2}}{2\sigma_{1}^{2}}})+
\frac{1}{2\pi\sigma_{2}^{2}}exp({-\frac{k_{T2}^{2}}{2\sigma_{2}^{2}}})
\end{equation}

\hspace{-0.4cm}This distribution was adding in PYTHIA code and calculated the
azimuthal correlations. The Figure 3 show the experimental data and the 
simulation. The simulation is in good agreement with the experimental data. 
The values of $<k_{T1}>$ and $<k_{T2}>$ are 0.558 $\pm$ 0.042 and 
$<k_{T2}>$ = 0.099 $\pm$ 0.050 respectively. In addition the magnitude of 
the partonic transverse momentum $<j_{Tz}>$ was
 calculated. The values obtained of $<j_{Tz}>$ = 0.397 $\pm$ 0.091  GeV/c are in
 agreement with the average value  $<j_{Tz}>$ = 0.324 $\pm$ 0.06 GeV/c obtained
 experimentally \cite{Rak:2004gk}.

\begin{figure}
\includegraphics[height=.4\textheight]{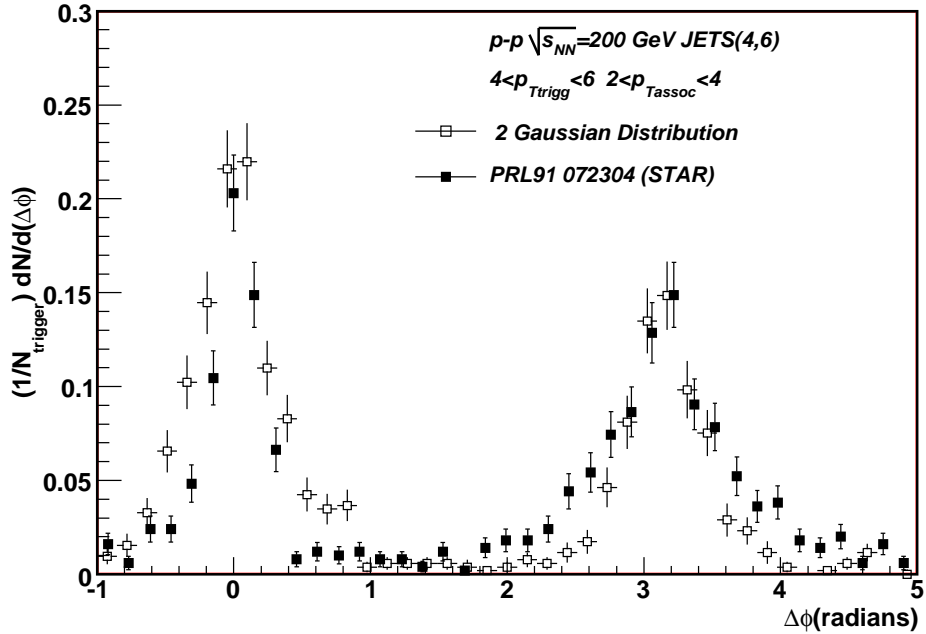}
\caption{Azimuthal distributions for p+p collisions, experimental data (line)
  \cite{Adams:2003im} and  simulations with two Gaussian distributions for the partons intrinsic
    momentum (circle)}
\end{figure}

\section{Summary}

We report the analysis of experimental azimuthal correlations measured by STAR
in p-p collisions at $\sqrt{s_{NN}}$ = 200 GeV. Comparisons between
experimental data and simulation with different $<k_{T}^{2}>$ show that the
$<k_{T}^{2}>$ characterized by a Gaussian distribution can not
reproduce experimental data.

Assuming two Gaussians distributions for $k_{T}$ the simulation 
is in agreement with the experimental data, as far as, we understand the use of
two Gaussians. It has never been used before to explain the peaks observed in
azimuthal correlations.

In addition the magnitude of the partonic transverse momentum $<j_{Tz}>$ was
 calculated. The values of $<j_{Tz}>$ =  0.397 $\pm$ 0.091 GeV/c are in
 agreement with the average value  $<j_{Tz}>$ = 0.324 $\pm$ 0.06 GeV/c obtained
 experimentally.

\begin{theacknowledgments}
The authors thanks A. Morsch for his valuable comments and
suggestions. Support for this work has been received by PAPIT-UNAM under grant
number IN107105.
\end{theacknowledgments}

\end{document}